\documentclass[fleqn,usenatbib]{mnras}

\usepackage{newtxtext,newtxmath}
\usepackage{comment}
\usepackage[T1]{fontenc}

\DeclareRobustCommand{\VAN}[3]{#2}
\let\VANthebibliography\thebibliography
\def\thebibliography{\DeclareRobustCommand{\VAN}[3]{##3}\VANthebibliography}

\usepackage[switch]{lineno}
\usepackage{graphicx}	
\usepackage{amsmath}	
\usepackage{orcidlink}
\usepackage{siunitx}

\title[AstroSat UV Deep Field IV: XUV at z=0.67]{AstroSat UV Deep Field IV. An Extended UV disk around a massive spiral galaxy at z=0.67}

\author[Pushpak Pandey et al.]{
Pushpak Pandey\orcidlink{0009-0009-7497-3431},$^{1}$\thanks{E-mail: pushpak@iucaa.in}
Kanak Saha\orcidlink{0000-0002-8768-9298},$^{1}$
Sanchayeeta Borthakur\orcidlink{0000-0002-2724-8298}$^{2}$
\\
$^{1}$Inter-University Centre for Astronomy and Astrophysics, Ganeshkhind, Post Bag 4, Pune 411007, India\\
$^{2}$School of Earth and Space Exploration, Arizona State University, 781 E Terrace Mall, Tempe, AZ 85287-1404, USA
}

\date{Accepted XXX. Received YYY; in original form ZZZ}

\pubyear{\the\year{}}

\begin{document}
\label{firstpage}
\pagerange{\pageref{firstpage}--\pageref{lastpage}}
\maketitle

\begin{abstract}
Extended ultraviolet (XUV) emission in nearby disk galaxies supports the inside-out growth scenario through low-efficiency star formation in their outer regions. However, such detections have largely been limited to the local Universe ($z \sim 0$) due to the need for deep, high-resolution UV imaging. We report the detection of a clumpy XUV disk in a massive, isolated spiral galaxy ($\log(M_*/M_\odot) \approx 11.04$) at $z=0.67$, observed with AstroSat/UVIT. The intrinsic rest frame FUV surface brightness profile, corrected for the instrument PSF, shows a more extended disk than its optical and IR counterparts. The XUV disk reaches nearly twice the optical radius and includes a large UV-bright low surface brightness (LSB) region ($S_{LSB}/S_{K80}\approx15,\ \mu_{FUV}-\mu_K\approx0.8$), consistent with the Type II XUV definition. Additionally, the detection of UV clumps without optical counterparts supports a Type I classification, suggesting gravitational instabilities and recent star formation. These features point to recent cold gas accretion onto the outer disk. From the asymmetric light profile, we estimate a gas accretion rate of $\sim11\ M_\odot$ yr$^{-1}$, providing evidence of active disk growth at intermediate redshift.
\end{abstract}

\begin{keywords}
galaxies: star formation -- Galaxy: disc -- Galaxy: evolution -- galaxies: spiral
\end{keywords}

\section{Introduction} \label{sec:intro}
The process of galaxy growth and evolution can be directly attributed to the process of star formation, as it can be triggered by various phenomena such as disk instabilities, mergers, gas accretion \citep{2009Dekel_2009_Gal_form,2008Hopkins_merger,2012Kennikut_Evans_SF,2015Sommerville}, etc. Hence, star formation can act as a reliable tracer for such events.
One such mechanism that may cause recent star formation in the outer disk region is the infall of cold gas from the intergalactic medium (IGM), which can form star-forming clumps around the disk and result in extended ultraviolet (XUV) emission beyond the conventional star formation threshold. This phenomenon was first observed in galaxies such as NGC 4625 \citep{Gil_de_paz_2005} and M83 \citep{Thilker_2005ApJ...619L..79T}.

Galaxies with such extended UV emission beyond the star formation threshold surface brightness ($\mu_\mathrm{FUV} = 27.25$ mag arcsec$^{-2}$ or $\Sigma_\mathrm{SFR} = 3 \times 10^{-4}\ M_\odot\ \mathrm{yr}^{-1}\ \mathrm{kpc}^{-2}$), where Kennicutt-Schmidt law breaks down, are termed XUV disks. According to \citet{Thilker_2007ApJS..173..538T} (T07), such systems are classified as Type I XUV disks if they show discrete UV-bright clumps in the outer disk, and Type II if the star formation is more diffuse, covering a low surface brightness (LSB) zone beyond the $K_{80}$ radius, the radius enclosing 80\% of the rest-frame $K$-band light, and within $R_{SFR}$, the radius where restframe FUV magnitude is equal to the SFR threshold ($\mu_{FUV}=27.25$). These definitions have since been adopted, to varying degrees, in follow-up studies \citep{Lemoinas2011, Moffett2012, Mousumi_Das_XUV}. Other works have also used the $R_{25}$ radius \citep{Zaritsky2007, Goddard2010} or break radius in surface brightness (SB) profiles \citep{Pohlen_Trujillo2006A&A...454..759P, Suchira2024} to define the XUV region \citep{DIISC_II}.

The extent of XUV emission in such galaxies often extends 2–4 times beyond their optical extents and is directly correlated with the distribution of HI gas, which is known to reach much farther out than the stellar disk \citep{Thilker_2005ApJ...619L..79T, Zaritsky2007, DIISC_III}. Extended star formation in these outer regions can be driven by several processes, such as gas accretion from the IGM (often inferred from low metallicities; \citealt{Gil_de_paz_2007_metal, Goddard_2011}), tidal interactions \citep{Neff_Tidal, Lopez_Tidal_XUV}, spiral density waves, or even radiative cooling \citep{Dekel_Radiative_cooling}. While galactic winds may redistribute gas within the disk, they are generally considered secondary contributors \citep{Thilker_2007ApJS..173..538T}.

From the perspective of galaxy evolution, the detection of XUV emission supports the inside-out disk growth paradigm \citep{Thilker_2007ApJS..173..538T,2007Muenoz_Gil,2022NaturBorgohain,DIISC_III} and allows us to trace how galaxies build up their outer disks over time. Moreover, these XUV zones, particularly when forming stars at high sSFR ($\log \mathrm{sSFR} > -9$) and low stellar mass surface densities, can be considered localised starburst regions. Such rapid and concentrated star formation may have consequences on the interstellar medium (ISM). However, given the low stellar densities and the likely absence of sustained supernova (SN) feedback in these outer regions, the ISM properties in the XUV zone may be markedly different from the more evolved ISM of the central disk.

XUV disks are found in $\sim 20$–30\% of nearby disk galaxies \citep{Thilker_2007ApJS..173..538T, Gil_de_paz_2007, Lemoinas2011}. While the phenomenon is often associated with spiral galaxies, extended UV emission is also observed in some blue early-type systems \citep{Moffett2012, Dhiwar_Lstar}, which may be rebuilding their disks. Due to the diversity in morphology, these galaxies span a wide region in the colour–magnitude diagram. The detected number density of XUV disks appears relatively uniform across stellar mass bins, but their relative fraction is higher in the high-mass regime \citep{Lemoinas2011}.

The typical surface brightness of XUV features in galaxies is $\mu (mag/arcsec^2)\gtrsim 26$, and 
detecting such XUV features at higher redshifts is observationally challenging due to their low surface brightness and the need for sub-kpc resolution. GALEX-based studies have been largely limited to the nearby Universe ($z \lesssim 0.05$). A rare example of intermediate-redshift XUV detection is provided by \citet{2022NaturBorgohain} (B22), using UVIT data to identify extended UV emission in blue compact dwarfs at $z \sim 0.1$–0.24 using PSF-corrected SB profile fitting. In this work, we present the discovery of an XUV disk in a massive spiral galaxy at $z \approx 0.67$, observed via the N242W band of UVIT onboard AstroSat. This is the most distant and most massive star-forming spiral galaxy hosting an XUV disk to date.

In the following sections, we describe our data, and the disk modeling process (similar to B22) used to estimate the UV star formation threshold radius and $K_{80}$. We further estimate the gas accretion rate from lopsidedness in the stellar disk, and discuss this system's properties in comparison with literature XUV samples to build a case for its inside-out disk growth

\section{Data} 
\label{sec:data}

\begin{figure*}
    \includegraphics[width=1.0\textwidth]{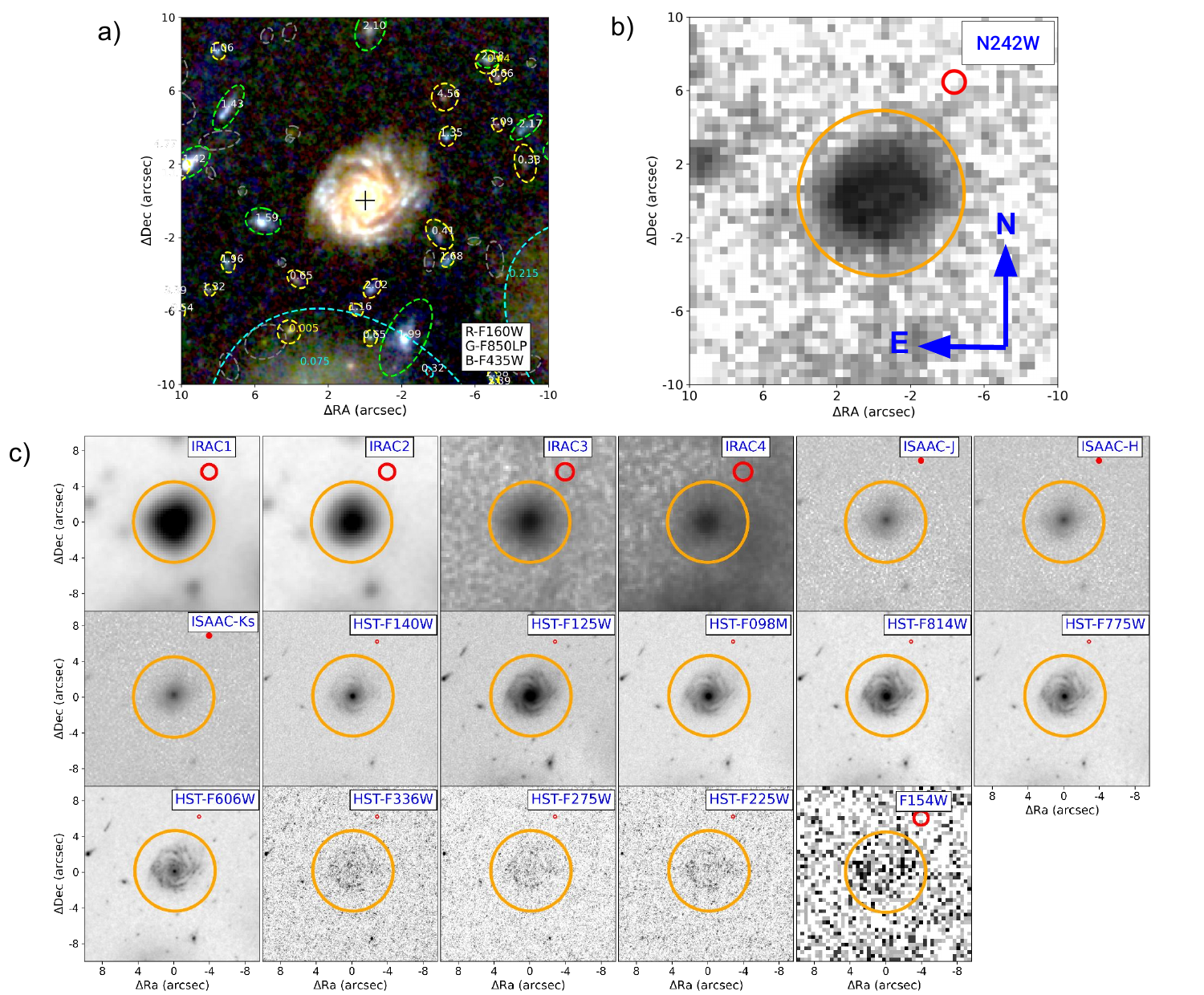}
    \caption{
\textbf{(a)} RGB image of AUDFs\_N02766 constructed from HST bands F160W (red), F850LP (green), and F435W (blue). Nearby HST detections from the HLF catalog are marked with dashed ellipses, with redshift types color-coded as follows: \textcolor{cyan}{cyan} (spectroscopic), \textcolor{green}{green} (grism), \textcolor{yellow}{yellow} (photometric), and \textcolor{gray}{gray} (unknown).
\textbf{(b)} UVIT/N242W image of AUDFs\_N02766, probing the rest-frame FUV ($\sim1500\,\text{\AA}$) used for XUV detection.
\textbf{(c)} Multiwavelength imaging from UVIT, HST, VLT/ISAAC, and Spitzer/IRAC, spanning $\sim1300\,\text{\AA}$ to $9.5\,\mu$m. Red circles below the filter labels indicate the PSF FWHM of each band. The orange circular aperture ($r \sim 4.58\arcsec$), centered on the galaxy, is used for photometry in the SED modeling.
}

\label{fig:rgb_XUV_AUDF-X01}
\end{figure*}

The galaxy AUDFs\_N02766 (\citealt{kanak_saha_AUDFCat}, also cataloged as GOODS-S 38092; \citealt{SWM_2014}), shown in Figure~\ref{fig:rgb_XUV_AUDF-X01}a, is located at ICRS coordinates RA = $53.1251^{\circ}$, Dec = $-27.7299^{\circ}$ (J2000), with a redshift of $z = 0.668$, corresponding to a lookback time of approximately 6.12 Gyr. It is a massive, star-forming disk galaxy with prominent spiral arms, clumps, and a significant bulge visible in UV, optical and infrared bands, see Figure~\ref{fig:rgb_XUV_AUDF-X01}. In HST imaging, it spans $\sim$9$\arcsec$ along the semi-major axis - approximately 63 kpc, assuming a flat $\Lambda$CDM cosmology with $H_0 = 70$ km/s/Mpc and $\Omega_m = 0.3$ (used throughout this work).

We utilize archival GOODS-South observations from HST/ACS and WFC3 \citep{Windhorst_rogier_uv,Illingworth_HLF_optical,Whittaker_2019ApJS..244...16W}, VLT/ISAAC \citep{ISAAC_Goods_S_2010A&A...511A..50R}, and Spitzer/IRAC \citep{Greats_Stefanon_2021ApJS..257...68S}, as well as UVIT-AstroSat data \citep{kanak_saha_AUDFCat}. The galaxy’s redshift is further confirmed through its optical spectrum from VLT/VIMOS \citep{Spectrum_VLT_2010A&A...512A..12B}, classified as Quality ‘A’ with clearly detected features including the Balmer 4000 Å break, [O III] doublet, and Ca II H $\&$ K absorption lines (see appendix \ref{appendix:Spectrum_and_obs}).

\section{Background subtraction}

To estimate the local background in the UVIT data, we make use of the background maps of the GOODS-South field from \citet{kanak_saha_AUDFCat}, constructed following the method described in \citet{Minima_stat}. These maps provide an initial background subtraction for the UVIT images. To further remove any residual background in the vicinity of our source, we run \textsc{SExtractor} on the N242W cutout to generate a segmentation map. We then place fifty $5\times5$ pixel boxes ($\sim$4.35 arcsec$^2$) within the unsegmented regions and compute the mean background from these apertures. This residual background estimate is subsequently subtracted from the UVIT cutouts. A similar procedure is followed for the HST images. Residual background levels are estimated and subtracted using the segmentation maps from \citet{Whittaker_2019ApJS..244...16W}.

\begin{figure}
    \centering
    \includegraphics[width=1\columnwidth]{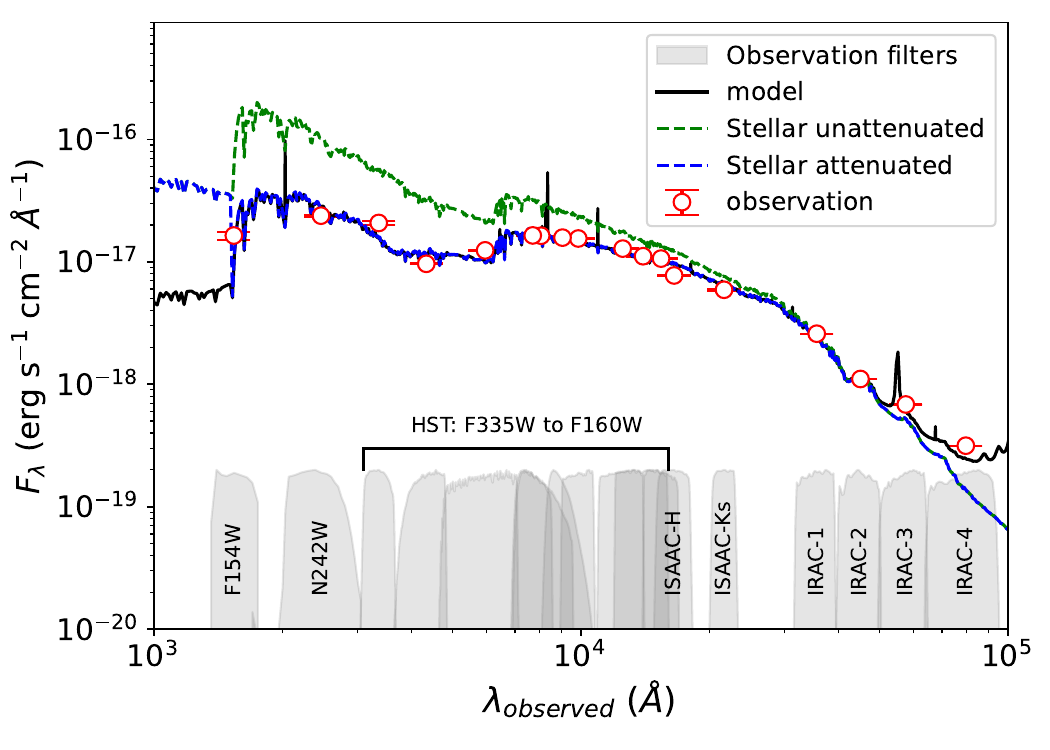}   \caption{The best-fit SED model (black) is shown alongside the observed fluxes (red points). The attenuated and unattenuated stellar continua are shown as blue and green dashed lines, respectively. The reduced $\chi^2$ for the fit is 1.2. The transmission curves (scaled) of the filters used in the fitting are overlaid as grey shaded regions.}
    \label{fig:SED}
\end{figure}

\section{Photometry and SED fitting}
\label{sec:sed}
To enable consistent flux measurements across multiple filters (Figure \ref{fig:rgb_XUV_AUDF-X01}) from different telescopes, the HST and ISAAC images are matched to UVIT/N242W PSF using the Photutils (details in appendix \ref{appendix:psf_match}) \citep{Photutils_2022zndo....596036B}. Since the F154W band is already closely matched to N242W beyond $r > 2\arcsec$, it requires no further adjustment \citep{kanak_saha_AUDFCat}. Due to the broad PSF of IRAC-2, 3, and 4 channels, we use aperture-corrected fluxes from the original images to avoid contamination from neighbouring sources. A common 4.58$\arcsec$ radius circular aperture was applied across all PSF-convolved images and their associated weight maps, centered on the bulge of the galaxy, to estimate fluxes and their uncertainties. Background and Poisson errors are calculated for the UVIT bands following \cite{Minima_stat}. Aperture corrected fluxes (converted to mJy and corrected for foreground extinction using \citealt{Foreground_ext}) are then used to construct the broadband SED ($1300\si{\angstrom} -  9.5 \mu m$) for the galaxy. We fit the SED using \textsc{CIGALE} \citep{Cigale} employing double exponential SFH, BC03 stellar population models, and Milky Way-like dust attenuation law. The best-fitting SED (reduced $\chi^2 = 1.2$, Figure \ref{fig:SED}) gives a total stellar mass of $1.08 \times 10^{11}\ M_\odot$, a young stellar mass of $2.54 \times 10^9\ M_\odot$ (100 Myr), a star formation rate of $28.1 \pm 1.4\ M_\odot,\text{yr}^{-1}$, and $E(B-V)=0.132$. For more details on PSF matching, photometry and SED modelling, see appendix \ref{appendix:psf_match}.

\section{Modelling of radial disk Profiles}
\label{sec:radial}

\begin{figure*}
    \includegraphics[width=1.0\textwidth]{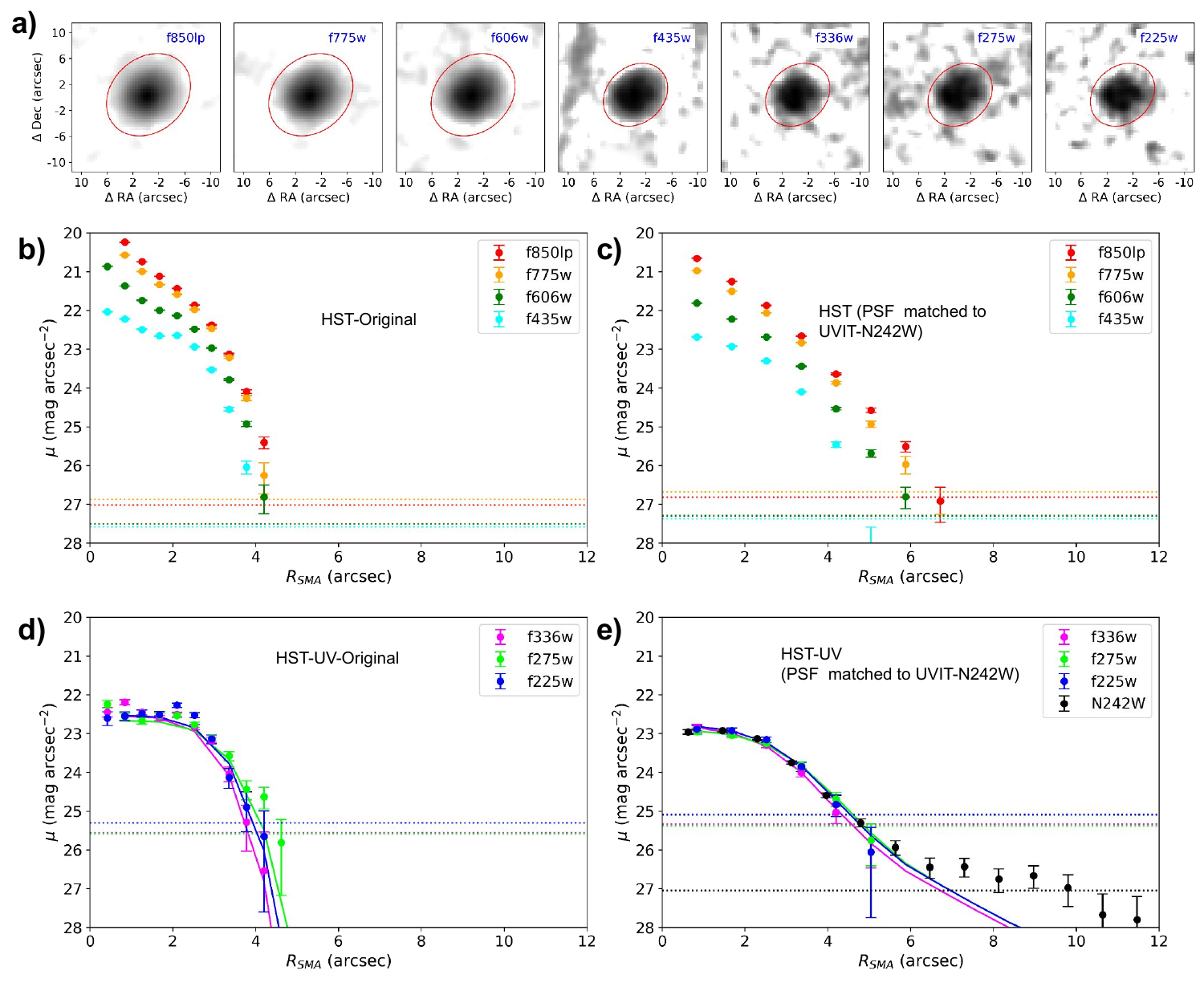}
    \caption{\textbf{(a)} UVIT-N242W PSF convolved images of AUDFs\_N02766 using cleaned HST images. The red ellipse represents the 3 SNR radius of the SB profile measured from these images. 
    \textbf{(b)} SB profiles from HST optical bands.
    \textbf{(c)} SB profiles from HST optical bands, convolved to UVIT-N242W psf. 
    \textbf{(d)} SB profiles from HST UV bands. 
    \textbf{(e)} SB profiles from HST UV bands, convolved to UVIT-N242W psf. The solid curves in (d) and (e) represent the sersic model used to fit the HST UV SB profiles upto their $3\sigma$ SB limit. The $3\sigma$ SB limit for this galaxy in each HST band are represented by the dotted horizontal lines}
    \label{fig:HST_compare}
\end{figure*}
In order to extract the radial surface brightness (SB) profile of the galaxy, we first estimate the disk ellipticity ($\epsilon$) and position angle (PA) using the HST/F160W image (Figure \hyperref[fig:rgb_XUV_AUDF-X01]{1a}). This is done by fitting isophotes with the Ellipse fitting routine, keeping the center fixed at the centroid of the bulge. While spiral arms in the inner region ($R < 3.5\arcsec$) influence the isophotal shape, the outer isophotes are relatively stable and yield a mean ellipticity of $\epsilon = 0.2$ and PA of $41.8^\circ \pm 0.8^\circ$. Using these parameters, we construct a set of elliptical annuli centered on the bulge and applied to all the cutout images in the optical/IR filters to extract the radial profiles and corresponding errors.
The rest-frame FUV SB profile has been extracted from the observed N242W filter image, to which we first mask the foreground/background sources using the \citet{Whittaker_2019ApJS..244...16W} segmentation map, convolved with a top-hat kernel of $2\times$ the N242W PSF FWHM (see Appendix \ref{Appendix:masking}).

All surface brightness (SB) profiles are corrected for cosmological surface brightness dimming \citep{Tolman1930, Tolman1934} due to redshift, yielding a correction of $A_{\rm cosmo} = -1.67$. We further apply a correction for internal dust attenuation using the \citet{Calzetti2000} extinction law, assuming a uniform colour excess $E(B-V)$ derived from our best-fit SED. This results in a correction of $A_{\rm dust, FUV} = -1.227$ for the rest-frame far-UV (corresponding to the N242W band). For more details on Surface Brightness correction, see Appendix \ref{Appendix:Surf_bright}. Additionally, we extract and present SB profiles for the PSF-convolved HST optical and UV bands matched to the UVIT N242W resolution (Figure~\ref{fig:HST_compare}), alongside their corresponding original HST SB profiles.

We consider a forward modelling approach to derive the intrinsic SB profile of the galaxy in a given passband. In this process, we first convolve a presumed model (say, exponential or Sersic profile) with a PSF and then fit the convolved profile with observed SB data to extract the best-fit model parameters. We employ the astropy.modeling framework \citep{Astropy_2022ApJ...935..167A} to generate two-dimensional exponential disk and Sersic models at the same pixel scale as the observed images. These models are convolved with the respective PSF and sampled using the same elliptical annuli (used to extract observed SB profile) to produce model 1D profiles.
We then fit these to the observed galaxy profiles using the least-squares routine from scipy.optimize \citep{2020SciPy-NMeth}.
A direct approach with 2D modelling using GALFIT \citep{2010Peng_Galfit} was tested but due to the SNR of individual pixels being low, we couldn't reach a reasonable fitting for the XUV region.
In most HST images, we find that beyond a radius of $\sim 4.5\arcsec$, the profiles show negative flux values. This indicates over-subtraction of the local background and results in a sharp decline in the radial profile near $\sim 4.5\arcsec$ (Figure \ref{fig:HST_compare}b). 

The HST UV SB profiles shown in Figure~\ref{fig:HST_compare}d,e exhibit a flat core within $\sim3\arcsec$, which is well described by a Sérsic profile up to the $3\sigma$ limit. When these best-fit models are extrapolated to larger radii ($\gtrsim4.5''$), and compared to the observed N242W SB profile, a clear UV excess emerges in the outer part of the galaxy.

The intrinsic scale lengths derived from best-fit model in various bands are summarised in Table~\ref{tab:scalelengths}. There seems to be a mild trend of increasing disk scale length at bluer filters.

\begin{table*}
\centering
\begin{tabular}{lllllll}
\hline
\textbf{Filter}  & \textbf{Restframe $\lambda$} & \textbf{Disk R$_s$(Kpc)} & \textbf{mag$_{\text{O}}$} & \textbf{SNR} & \textbf{Remarks} \\ \hline
& & &Global Properties\\
\hline
Irac-1                & 2200 nm                            & $6.3 \pm 0.08$ & $18.80\pm0.001$ &   -- &--                  \\ 
Isaac $K_s$            & 1320 nm                           & $6.1 \pm 0.14$         & $18.98\pm 0.02$ &   -- &--            \\ 
HST-F160W             & 960 nm                           & $6.5\pm0.14$   & $19.09\pm0.002$ &   -- &--             \\ 
HST-F140W             & 840 nm                           & $7.1 \pm 0.09$     & $19.34\pm0.006$ &   -- &--                 \\ 
HST-F775W             & 464 nm                           & $7.4\pm0.5$    & $20.12\pm0.004$ &   -- &--    \\
HST-F660W             & 363 nm                             & $6.5\pm0.5$     & $20.98\pm0.005$ &   -- &--                   \\ 
N242W              & 145 nm                             & $36- 6/+8$(XUV)      & $22.19\pm0.02$ &   -- &--                 \\ \hline
& & &UV clump properties\\ \hline
N242W (C1) & 145 nm & -- & $27.16 \pm {0.25}$ & 4.3  & - 
                       \\
N242W (C2) & 145 nm & -- & $27.32\pm{0.27}$ & 3.76 & - 
                      \\
N242W (C3) & 145 nm & -- & $27.19\pm{0.26}$ & 4.17 & - 
                         \\
N242W (C4) & 145 nm & -- & $27.53\pm{0.35}$ & 3.16 & - 
                         \\
N242W (C5) & 145 nm & -- & $27.49\pm{0.34}$ & 3.27 & - 
                       \\
N242W (C6) & 145 nm & -- & $27.07\pm{0.24}$ & 4.6  & phot-z $\sim$0.65 \textbf{$^*$}  \\
N242W (E1) & 145 nm & -- & $27.00\pm{0.20}$                & 4.84       & phot-z $\sim$4.56\textbf{$^{\dagger}$} \\ \hline

\end{tabular}
\caption{Global Properties of AUDFs\_N02766, along with detected UV clump properties.
Observed magnitudes in the filters (mag$_\text{O}$) and Disk scalelengths ($R_s$) are shown for the galaxy in separate filters. 
For clumps,shown in Figure \hyperref[fig:rgb_XUV_AUDF-X01]{1b} , we show mag$_\text{O}$ (in PSF FWHM sized aperture)
and signal-to-noise ratios (SNR). Possible contamination by nearby sources from the HLF catalog (within a 1.2$\arcsec$ aperture radius) is noted. Notes: $*$ The phot-z estimate is similar to  AUDFs\_N02766. ${\dagger} $Phot-z estimate of 4.56 places the N242W band in the Lyman continuum, where the transmission probability is very low \citep{Escape_fraction_Akio,KanakNature2020}.
}
\label{tab:scalelengths}
\label{tab:my-table_clump}
\end{table*}

\section{The clumpy XUV Disk}
\label{sec:XUV}

\begin{figure*}
    \centering
    \includegraphics[width=1\textwidth]{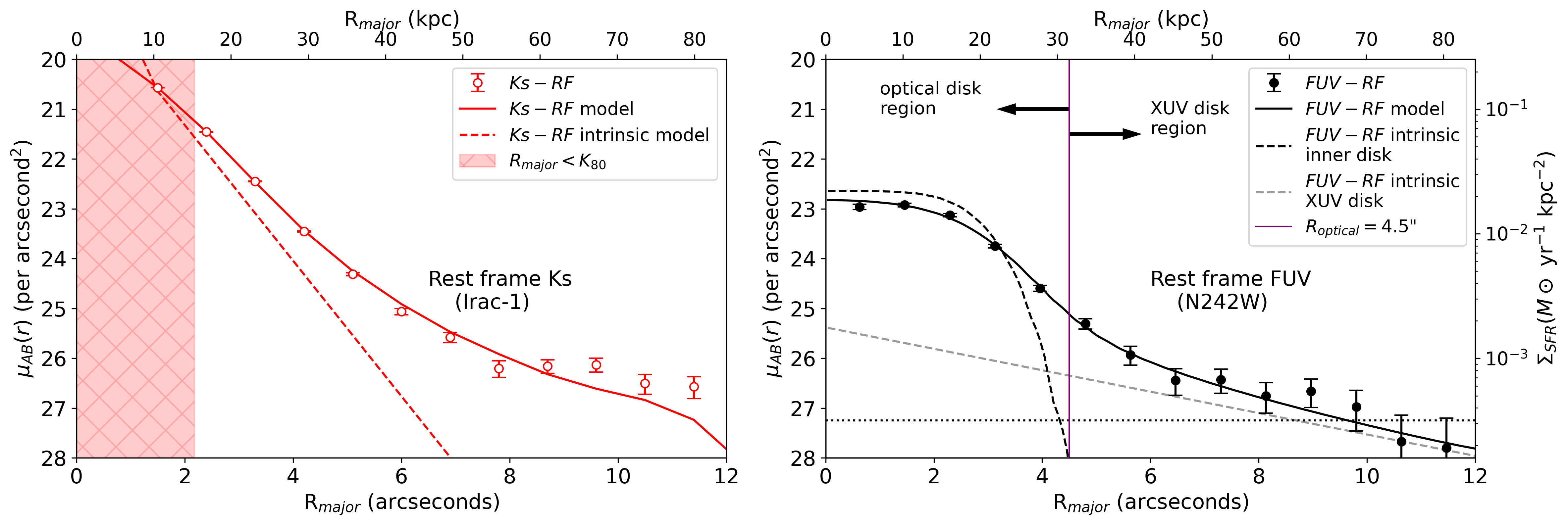}   
    \caption{Surface brightness profiles in rest frame Ks(IRAC-1, Left panel), rest frame FUV (UVIT-N242W, Right panel), corrected for cosmological dimming and extinction. PSF-convolved models (solid), intrinsic models (dashed), and sersic+ exponential for N242W (faint dashed) are shown. $K_{80}$ marks 80\% light radius of rest frame Ks (IRAC-1) model; $R_{SFR}$ (brown marker) marks $\mu_{FUV}=27.25$; magenta marker shows optical edge at $4.5\arcsec$(31.5 kpc), and pink marker marks the end of our XUV-detection boundary at $\sim10\arcsec $ ($\sim$70 kpc)}
    \label{fig:Uv_SB_Fitting}
\end{figure*}

Here, we follow the criteria outlined by T07, to characterise the extended ultraviolet (XUV) disk of AUDFs\_N02766. First, the intrinsic radial SB profiles of AUDFs\_N02766 in the rest-frame FUV and $K_s$ bands (IRAC-1) are being compared. A surface density threshold of $3 \times 10^{-4}\ M_\odot\ \mathrm{yr}^{-1}\ \mathrm{kpc}^{-2}$ - equivalent to $\mu_{\mathrm{FUV}} = 27.25$ mag arcsec$^{-2}$ - is adopted to identify the SFR limit in the rest frame FUV profile.
 
The radius encompassing $80\%$ of the total intrinsic restframe Ks band flux is $K_{80} = 2.18\arcsec$, while the rest frame FUV disk extends out to $R_{\mathrm{SFR}} \approx 8.6\arcsec$, where the corrected rest-frame FUV surface brightness reaches the SFR threshold (see Figure \ref{fig:Uv_SB_Fitting}). The low surface brightness (LSB) zone, defined between $K_{80}$ and $R_{\mathrm{SFR}}$, exhibits a color difference of $\mu_{\mathrm{FUV}} - \mu_{K_s} = 0.8$, and spans an area $\sim14.9$ times larger than the inner $K_{80}$ region. These criteria are consistent with a Type II XUV disk as per T07, which required the area of LSB zone ($S_{LSB}$) to be at least 7 times the $K_{80}$ region ($S_{K80}$), and the color $\mu_{\mathrm{FUV}} - \mu_{K_s} < 4$.

The intrinsic rest-frame FUV SB profile shows a noticeable transition from the bright inner disk to the fainter extended component at $\sim4$–$4.5\arcsec$, which coincides with the outer optical edge observed in most HST bands (Figure \ref{fig:HST_compare}b). Beyond this radius, the 1D profiles in the HST optical images drop below zero, likely due to background over-subtraction in Archival data. This suggests that the faint, extended disk is no longer visible in the optical bands but remains detectable in N242W. 

To classify the XUV disk of AUDFs\_N02766 as Type I, we run SExtractor with high deblending on the N242W image (more in section \ref{sec:SExtParam}) and identify several faint, clumpy UV structures (with SNR $> 3$).
 
\begin{figure*}
    \centering
    \includegraphics[width=1\textwidth]{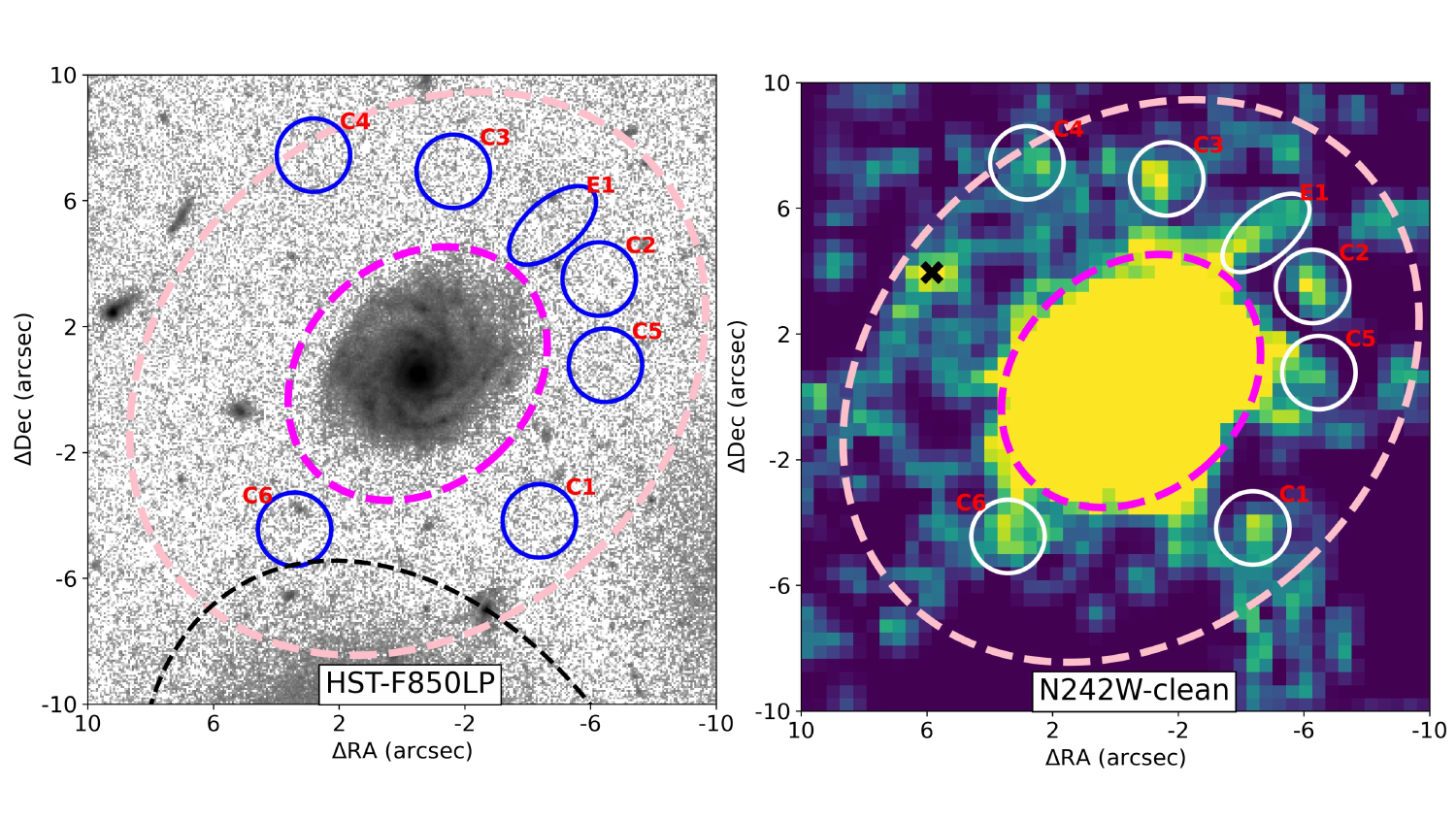}
    \caption{ \textbf{(a)}HST-F850LP band observation of AUDFs\_N02766, along with the blue apertures marking the clumps detected in N242W.
    \textbf{(b)} UVIT-N242W cleaned with mask prepared from HST Segmenation, showing the $>$ 3 SNR clump detections with white apertures.  The \textcolor{magenta}{magenta} ellipse denotes the optical edge ($R_{major}\sim4.5\arcsec$); the \textcolor{pink}{pink} ellipse outlines the radius where rest frame FUV SB profile (Figure \ref{fig:Uv_SB_Fitting}) drops to SNR$\approx$3 ($R_{major}\sim10\arcsec$). A residual bright feature in the N-E corner, left after masking the optical sources, is marked with a black cross, due to its clear association with a nearby source, and is excluded from consideration as an XUV clump.
    }
\label{fig:clumps}
\end{figure*}

During this process, we remove the UV clumps with optical counterparts within a PSF FWHM radius (1.2$\arcsec$). Many of these clumpy structures are identified at an $r>R_{SFR}$ which is consistent with the Type I XUV disk definition \citep{Thilker_2007ApJS..173..538T} (Figure \ref{fig:clumps}). The photometric details of each detected clump is provided in Table \ref{tab:my-table_clump}. 
A similar analysis with the archival HST/F225W and F275W data does not reveal these UV clumps, likely due to insufficient depth. We mark the XUV disk as the region beyond the $4.5\arcsec$ radius, which extends to $\sim10\arcsec$ in far-UV. Additional details of clump detections and clump properties are discussed in the supplementary material.

From the intrinsic surface brightness model of the rest-frame FUV, we calculate the integrated UV-based star formation rate (SFR) in the XUV region ($4.5\arcsec < R < 10\arcsec$) to be approximately $4.3\ M_{\odot}\ \mathrm{yr}^{-1}$ \citep{Kennicutt_SFR_1998ARA&A..36..189K}. Assuming a constant SFR over the past 100 Myr, this implies a young stellar mass of $\sim 4.3 \times 10^8\ M_{\odot}$ formed in this region, which is $\sim$15\% of the total young stellar population obtained via SED modelling in section \ref{sec:sed}.

\section{Robustness of clump detection and measurement}
\label{sec:SExtParam}

To detect clumps in the XUV region as individual sources, we adopt the hot-mode detection strategy of \citet{kanak_saha_AUDFCat} on the N242W band, with two modifications to \texttt{SExtractor} parameters: \texttt{DETECT\_MINAREA} = 5 pixels and \texttt{DEBLEND\_MINCONT} = $5\times10^{-6}$. These choices are motivated by our aim to identify extremely faint, compact clumps in the XUV disk. Since the clumps are expected to be marginally resolved, \texttt{DETECT\_MINAREA} is set to match the area subtended by the N242W PSF FWHM ($\sim5$–6 pixels), maximising sensitivity to faint point-like sources, while a lower \texttt{DEBLEND\_MINCONT} aids in separating overlapping clumps. All other parameters follow the standard hot-mode configuration. All optical HST sources within the XUV region appear point-like at the N242W resolution; to ensure the independence of detections, we search for optical counterparts within a $1.2\arcsec$ aperture (corresponding to the N242W PSF FWHM). Sources with clear HST counterparts at different redshifts are excluded, and for overlapping detections within a PSF FWHM radius, only the more significant source is retained. The remaining clumps without optical counterparts are listed in Table \ref{tab:scalelengths}.

Clump C6 has an optical counterpart with $z_{\rm phot}\approx0.65$, detected across all bands, and may represent either a satellite of AUDFs\_N02766 or a foreground/background galaxy. The elliptical feature E1 coincides with a source at $z_{\rm phot}=4.2$, for which the red edge of the N242W filter lies below the Lyman limit ($\lambda < 912,\text{\AA}$). Given the rarity of confirmed LyC emitters, we interpret E1 as a morphological feature associated with AUDFs\_N02766. We present a cleaned view of AUDFs\_N02766 after masking all nearby sources with optical counterparts (as identified in HST F850LP), except for the two sources overlapping with C6 and E1. In this image, all \textsc{SExtractor}-detected clumps are marked with white apertures in both HST/F850LP and N242W (Figure~\ref{fig:clumps}). A residual bright feature in the north-east corner, clearly associated with a nearby optical source, is excluded from further analysis.

\begin{figure*}
    \centering
    \includegraphics[width=1\textwidth]{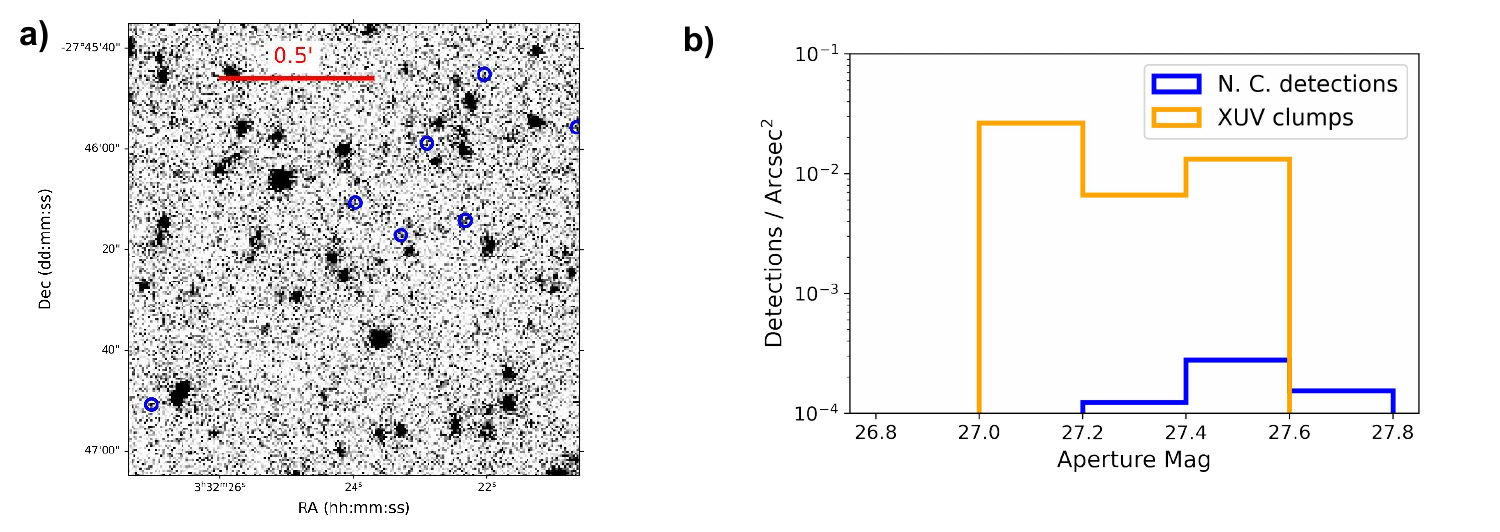}
    \caption{
    \textbf{(a)} A $1.5\arcmin \times 1.5\arcmin$ region within the $3\arcmin \times 3\arcmin$ N242W test patch in the AUDF GOODS--South field, located near AUDFs\_N02766. The blue apertures denote the $>3$ SNR detections with no HST-optical counterpart within $1.2''$ radius.
    \textbf{(b)} AB magnitude distributions: the orange histogram shows UV clumps density detected in the XUV region, while the blue histogram represents sources ($>$ 3 SNR) without HST counterparts (No Counterpart: N.C. detections), used to assess the density of probable spurious detections.
    }
    \label{fig:Clumps_tests}
\end{figure*}

\begin{figure*}
    \centering
    \includegraphics[width=1\textwidth]{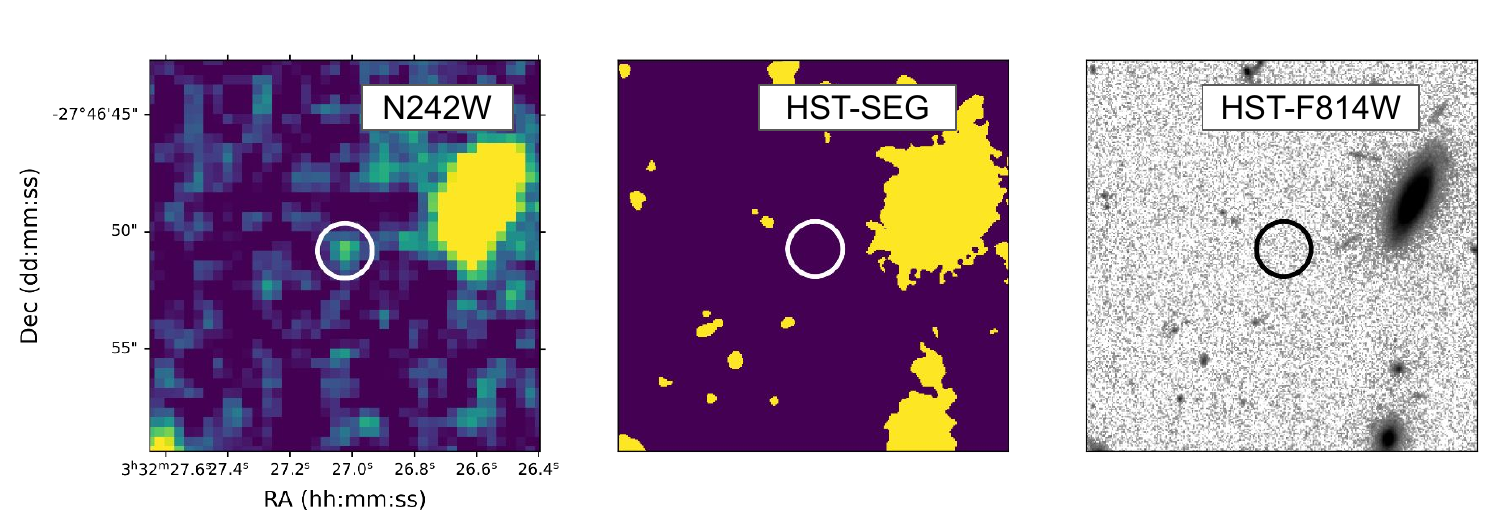}
    \caption{An example of N.C. (no-counterpart) sources identified in the first clump-detection robustness test after masking optical sources. The left panels show $20\arcsec \times 20\arcsec$ UVIT/N242W cutouts centered on the detection, the middle panels show the corresponding HST segmentation maps, and the right panels display HST/F814W images of the same regions. The absence of significant detections in the HST panels confirms the lack of plausible optical counterparts to these UV sources.}
    \label{fig:Spurious_sources}
\end{figure*}

\noindent
To assess the reliability of our detections and quantify spurious sources, defined as detections that might arise dues to noise peaks, detectable by our SExtractor parameters as legitimate $>$3 SNR sources, we perform two tests using a $3\arcmin \times 3\arcmin$ patch in the GOODS-South field (Figure~\ref{fig:Clumps_tests}), located $\sim5\arcmin$ south of AUDFs\_N02766, where background conditions are comparable. In the first test, we run \texttt{SExtractor} after masking optical sources using the HST segmentation map convolved with a tophat kernel of radius $\approx$ N242W PSF FWHM. This yields 18 detections (an example shown in Figure \ref{fig:Spurious_sources}) with magnitude $<27.73$ and no optical counterparts, corresponding to a noise peak expectation of $6\times10^{-4}$ sources arcsec$^{-2}$, implying an expectation of 0.09 spurious clumps over the XUV area ($\sim151$ arcsec$^2$). A comparison of the $>3\ SNR$ detections per square arcseconds along with the clump density in XUV region is shown in Figure \ref{fig:Clumps_tests}b.
In the second test, we generate Poisson noise realisations matching the N242W image properties and run \texttt{SExtractor} with identical parameters, detecting 16 sources with magnitude $<27.73$, corresponding to $5\times10^{-4}$ sources arcsec$^{-2}$ and an expectation of 0.08 clumps. In both cases, contamination from noise peaks is negligible.

\section{Lopsidedness and Gas Accretion Rate}

\begin{figure*}
    \centering
    \includegraphics[width=1\textwidth]{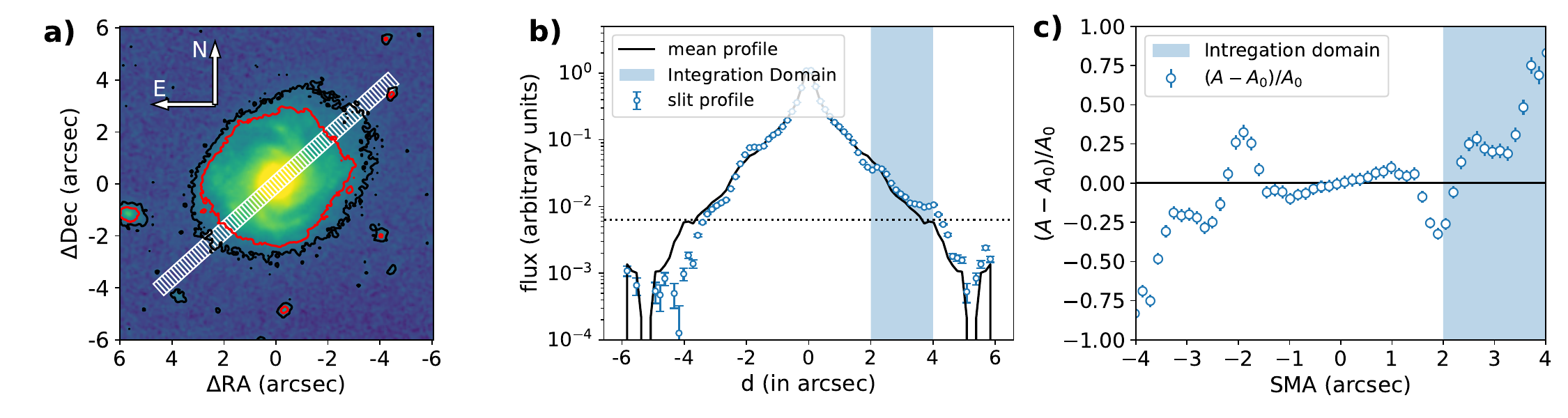}
    \caption{ \textbf{(a)} False-color image of stacked HST bands (F125W, F140W, F160W) with 1$\sigma$ (black) and 3$\sigma$ (red) isophotal contours overlaid. A white slit is placed along the direction of observed lopsidedness. \textbf{(b)} Light profile of AUDFs\_N02766 along the slit (blue markers) compared to the mean profile ($A_0$, black). The black horizontal dotted line represents the $1\sigma$ level shown in (a). \textbf{(c)} Radial variation of $A_1/A_0$. The mean $<A_1/A_0>$ is computed over the radial range highlighted in the blue region.  }
    \label{fig:lopsided}
\end{figure*}

Visual inspection of the HST images of AUDFs\_N02766 reveals a lopsided asymmetry in the light distribution along the galaxy’s semi-major axis, particularly toward the north-west, relative to the central bulge (Figure~\ref{fig:lopsided}). This asymmetry is subtle in the N242W image. Additionally, the N242W data show an elongated clumpy structure in the same direction (marked as E1 in Figure \hyperref[fig:rgb_XUV_AUDF-X01]{1b}). The prevalence of these asymmetric features help us understand gas inflow via angular momentum transport\citep{Sahajog2014}. While quantifying these asymmetries provides a useful proxy to estimate the galaxy’s recent gas accretion activity \citep{1997Rix_Zaritsky,JogCombes2009}. In order to estimate asymmetries in the galactic disk, we perform a Fourier decomposition of the light distribution:

\begin{equation}
A(r,\phi) = A_0(r)+\sum_{m=1}^{\infty}A_m(r)e^{in(\phi-\phi_m)},
\end{equation}

\noindent where $A(r,\phi)$ is the observed surface brightness, $A_0(r)$ is the azimuthally averaged radial profile, and $A_m$ and $\phi_m$ are the amplitude and phase of the $m$-th Fourier component. The $m=1$ mode quantifies lopsidedness, while higher-order terms ($m \geq 2$) capture finer asymmetries such as spiral arms or clumps \citep{Rix_Zaritsky_1995ApJ...447...82R}. To quantify lopsidedness, we follow the method of \citet{Rix_Zaritsky_1995ApJ...447...82R,1997Rix_Zaritsky}, estimating $A_1$ by fitting sinusoidal modes to the 2D light distribution $A(r,\phi)$. The HST/F160W intensity map is binned into a $25\times25$ grid in polar coordinates spanning $0 < r < 4.5\arcsec$, and $0 < \theta < 2\pi$. We compute the mean fractional lopsidedness $\langle A_1/A_0 \rangle$ in the radial range $2\arcsec < r < 4\arcsec$, where the $A_1$ phase angle is nearly flat. We find $\langle A_1/A_0 \rangle = 0.27 \pm 0.007$ in F160W, and $0.21 \pm 0.009$ in ISAAC-$K_s$. The phase angle of the $A_1$ mode is $64.9^\circ \pm 1.7^\circ$ in F160W and $67.6^\circ \pm 4.0^\circ$ in $K_s$, implying an offset of $\sim23-25^\circ$ from the Position angle of the galaxy.

To independently confirm this measurement, we co-add HST: F125W, F140W, and F160W images (PSF-matched and weighted by inverse variance), and extract a 1D light profile along a custom slit aligned with the galaxy’s position angle (see Figure \hyperref[fig:lopsided]{3a}). With the azimuth fixed, the light profile $A(r)$ can be decomposed as:
\begin{equation}
    A(r)\approx A_0(r) + A_1(r)
    \label{eqn:slit_decomp}
\end{equation}
where $r$ runs from $-L/2$ to $L/2$ along the slit, and $A_0(|r|) = \frac{1}{2} \left[A(-r) + A(r)\right]$. The lopsided component is computed as $A_1 = A - A_0$, and the relative amplitude $A_1/A_0$ is measured as a function of $r$ (Figure \hyperref[fig:lopsided]{3c}). Averaging over $2\arcsec < r < 4\arcsec$, we find $\langle A_1/A_0 \rangle = 0.260 \pm 0.002$, in agreement with the 2D Fourier result. According to \citet{1997Rix_Zaritsky}, a lopsidedness amplitude $\langle A_1/A_0 \rangle \geq 0.2$ is indicative of a galaxy that has accreted $\sim10\%$ of its mass within the past 1 Gyr. 
Assuming low merger activity (as the galaxy resides in a low density environment), for AUDFs\_N02766, this corresponds to an average gas accretion rate of $\sim11\ M_\odot,\mathrm{yr}^{-1}$. This rate is enough to replenish the depleting gas reservoirs in the XUV region, as the calculated UV SFR of $4.3\ M_{\odot}\ \mathrm{yr}^{-1}$ is less than the gas accretion rate calculated above.

\section{Discussion} \label{sec:discussion}
\subsection{Efficiency of UV disk models as a reliable UV SFR tracer}

A key limitation in determining the SFR threshold radius from UVIT observations arises from its point-spread function (PSF). The N242W band has a PSF FWHM of $1.2\arcsec$, corresponding to a physical scale of $\sim8.4$ kpc at $z = 0.67$, in contrast to the $\sim1$ kpc resolution of HST/F225W and F275W (\textbf{Figure \hyperref[fig:rgb_XUV_AUDF-X01]{1}}). Despite this lower spatial resolution, UVIT achieves a $3\sigma$ limiting magnitude of 27.73 mag (within a $1.2\arcsec$ aperture), deeper than the $\sim$25.9 mag limits of the archival HST UV data \citep{Windhorst_rogier_uv}. This makes UVIT more sensitive to diffuse and faint features such as extended UV (XUV) emission and disjoint clumps that remain undetected in HST UV bands.

However, the broader PSF in N242W leads to flux boosting in the galaxy’s outer regions, potentially biasing the inferred SFR threshold radius outward. To mitigate this effect, we apply a PSF-corrected disk-fitting algorithm to reconstruct the intrinsic radial surface brightness (SB) profile in the rest-frame FUV.

Our Type II XUV disk classification is based on modelling the rest-frame FUV profiles corrected for dust attenuation from the SED model. If measurements are done without any dust correction, the SFR limit radius reduces to $\sim4.23\arcsec$, decreasing the LSB region to $\sim3\times K_{80}$ region, and yielding a color of $\mu_{\mathrm{FUV}} - \mu_{Ks} \sim 2.26$. These values no longer satisfy the criteria for a Type II XUV disk as per \citet{Thilker_2007ApJS..173..538T}. Thus, without a radial attenuation model, which is not feasible with the current dataset, the Type II classification of AUDFs\_N02766 faces uncertainty. Nevertheless, the detection of multiple isolated UV-bright clumps and the faint exponential disk in the outer region (beyond the SFR threshold radius), without optical counterparts, supports the identification of AUDFs\_N02766 as a Type I XUV disk. 

\subsection{Star formation in XUV region and central quenching}

Comparison of the N242W observations with archival optical and infrared data reveals that star formation in AUDFs\_N02766 is spatially extended across the stellar disk. The dust-corrected, intrinsic rest-frame FUV SB profile shows that the SFR threshold lies well beyond the optical radius, whereas the older stellar population is concentrated near the central bulge. This central buildup may result from the cold gas accretion in the outer parts, forming clumps and their subsequent inward migration, potentially driving morphological transformation toward an early-type system \citep{Elmegreen_2008, 2009Dekel_clumpy, 2018Saha_S0,2022NaturBorgohain}. From our best-fit rest-frame FUV model, we estimate the total SFR in the XUV region ($4.5\arcsec < r < 10\arcsec$) to be $\sim4.3\ M_\odot\ \mathrm{yr}^{-1}$. Since the reported star formation efficiency of outer disk regions of galaxies, in general, is very low, the gas reserves would not be depleted by the ongoing star formation within a Hubble timescale \citep{2010Bigiel_SFE,2022NaturBorgohain}. The inferred gas accretion rate from the intergalactic medium is $\sim11\ M_\odot\ \mathrm{yr}^{-1}$ indicating a gas replenishment rate that exceeds ongoing star formation and is capable of sustaining the observed growth in the outer disk for a very long time. 

Within the inner disk ($r < 4.5\arcsec$), the intrinsic rest-frame FUV model-derived SFR is relatively high at $\sim25\ M_\odot\ \mathrm{yr}^{-1}$, despite the central rest-frame FUV surface brightness profile being notably flat. This may indicate the onset of central quenching \citep{2015Tacchella_Inside_Out_Quench}, consistent with the formation of a passive bulge and declining star formation in the core. The presence of an active galactic nucleus (AGN) remains a possibility, potentially contributing to gas removal or heating through feedback, thereby suppressing star formation in the central region \citep{2012Fabian_AGN_Quenching}.

\subsection{A comparison with other XUV studies }

\begin{figure}
    \centering
    \includegraphics[width=1.1\columnwidth]{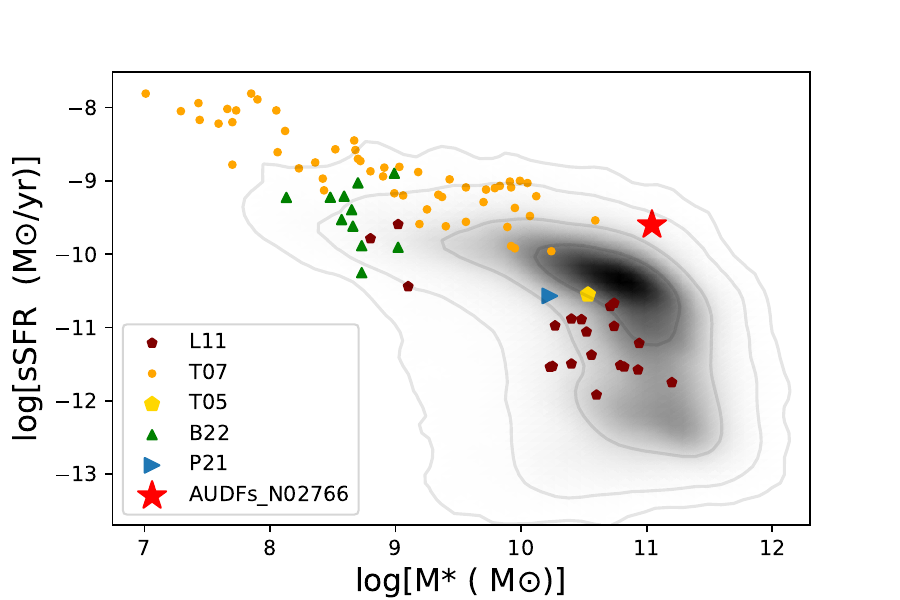}
    \caption{ A comparison of sSFR vs stellar mass of AUDFs\_N02766 and all the detected XUV disks (L11:\citet{Lemoinas2011}, T07:\citet{Thilker_2007ApJS..173..538T}, T05: \citet{Thilker_2005ApJ...619L..79T}, B22:\citet{2022NaturBorgohain}, P21:\citet{DIISC_II}) along with the population of galaxies at $z\approx 0$ shown in Black \citet{Salim_2016ApJS..227....2S}. }
    \label{fig:lit_surv2}
\end{figure}

In Figure~\ref{fig:lit_surv2}, the specific star formation rate (sSFR) of AUDFs\_N02766 is noticeably higher than that of the local star-forming galaxy population. A general anti-correlation is observed in the $\log$(sSFR) versus $\log$(stellar mass) plane for all detected XUV disks, relative to the average galaxy population.

AUDFs\_N02766, however, stands out with a significantly higher sSFR compared to locally identified massive XUV systems ($z \sim 0$). This enhancement may reflect the general increase in gas fraction with redshift \citep{Nissim_HI_2016ApJ...818L..28K}, which provides more fuel for star formation in the outer disk. Furthermore, the different environments and composition of galaxies at higher redshifts could play a critical role in shaping the properties and extent of XUV emission.

The anti-correlation in Figure~\ref{fig:lit_surv2} could be driven by the shallow potential wells of low-mass galaxies, where gas can be more diffusely distributed, enabling more extended star formation. Additionally, most photometric surveys rely on optical bands for source detection, which can miss faint, extended UV features in such galaxies. As a result, lower-mass XUV galaxies might be deriving a larger fraction of their total star formation from the XUV region. We may currently be missing such XUV disk galaxies at intermediate redshifts, which could be promising targets for future wide-field surveys such as the Vera C. Rubin Observatory \big(LSST; \citet{LSST_ivezi}\big), as well as upcoming UV missions \big(e.g. CASTOR;  \citet{CASTOR_UV_telescope}\big).

\section{Conclusions}
We summarize the key findings of this study below:
\begin{itemize}
    \item We report the detection of a very massive and strongly star-forming extended ultraviolet (XUV) disk galaxy, AUDFs\_N02766, at a redshift of $z \approx 0.67$. The XUV emission extends to nearly twice the optical size of the galaxy.
    \item  This represents the most distant XUV disk galaxy identified to date.
    
    \item  AUDFs\_N02766 exhibits a well-defined central bulge composed of older stellar populations, accompanied by widespread star formation across the disk and clumpy, UV-bright features in the outer regions. These characteristics support its classification as a mixed-type XUV disk (Types I + II).
    
    \item From its measured lopsidedness, we estimate an average gas accretion rate of $\sim 11\ M_\odot\ \mathrm{yr}^{-1}$, that appears sufficient to sustain the ongoing star formation in the XUV disk. 
    
\end{itemize}

\section*{Acknowledgements}
We thank the anonymous referee for their constructive feedback, which improved this manuscript. P.P. acknowledges Anshuman Borgohain, Soumil Maulick, Suraj Dhiwar, and the AUDF group for valuable discussions and input on the methodology.

\section*{Data Availability}
The UVIT data used in this study are not yet publicly available, but can be provided upon reasonable request to the corresponding author. All other datasets used in this work are publicly available from archival sources.

\bibliographystyle{mnras}
\bibliography{example} 

\appendix

\section{Spectrum}
\label{appendix:Spectrum_and_obs}
\begin{figure*}
    \centering
    \includegraphics[width=1\textwidth]{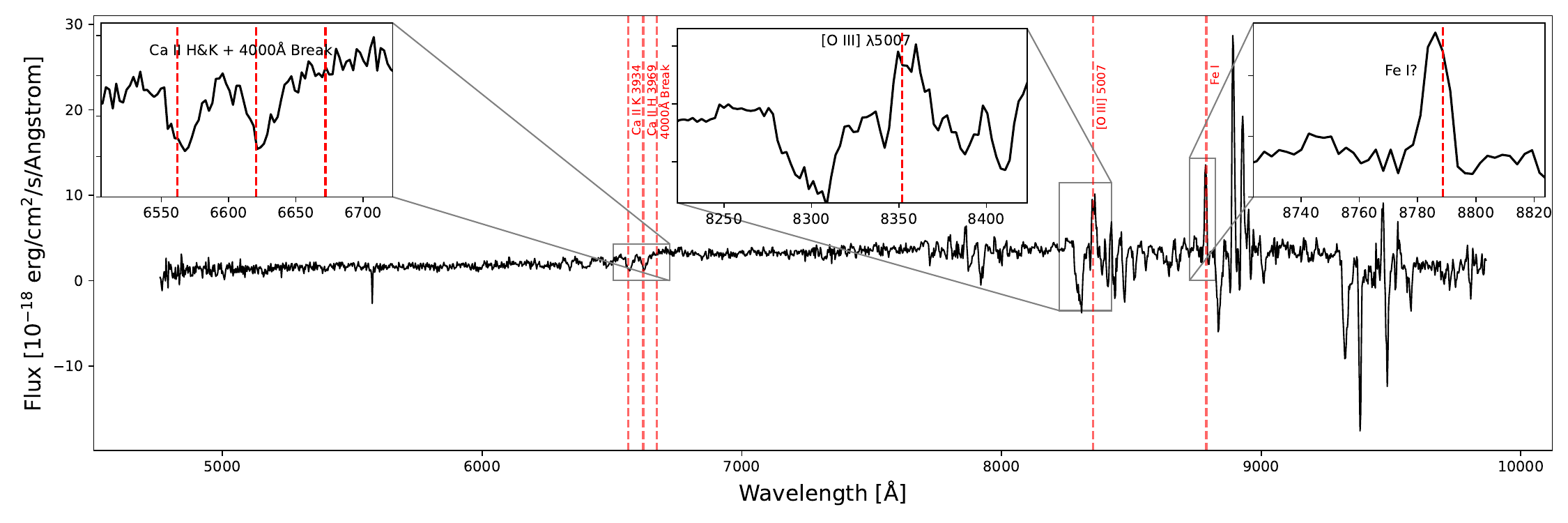}
    \caption{ The optical spectrum of AUDFs\_N02766 as observed by the VLT-Vimos survey \citep{Spectrum_VLT_2010A&A...512A..12B} }
    \label{fig:Spectra}
\end{figure*}

The released spectrum of AUDFs\_N02766 from VLT-Vimos survey by \citep{Spectrum_VLT_2010A&A...512A..12B} is shown in figure \ref{fig:Spectra} along with highlighted features, which were prominent and helped in redshift determination. 

\section{PSF matching, Photometry and SED modelling}
\label{appendix:psf_match}
\subsection*{PSF matching}
\begin{figure*}
    \centering
    \includegraphics[width=1\textwidth]{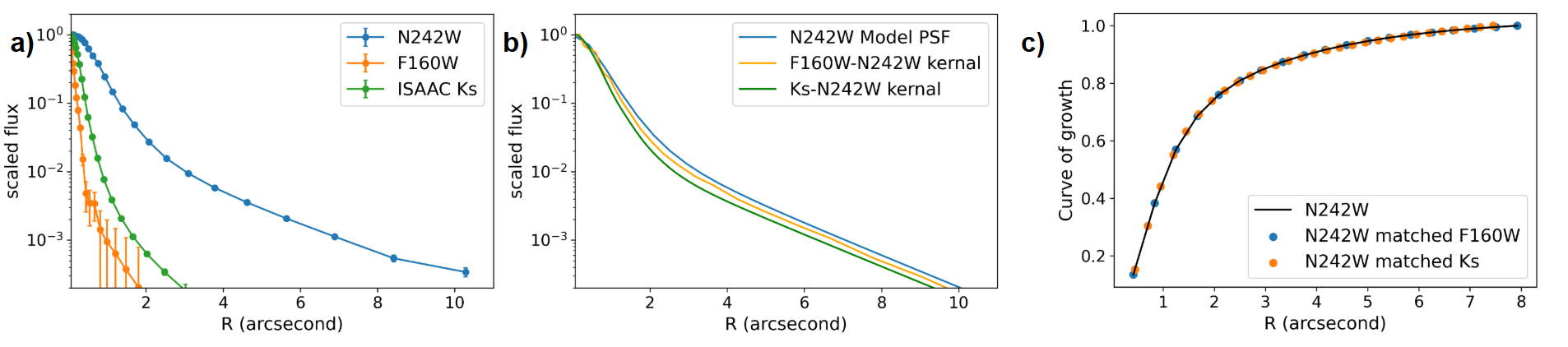}
    \caption{ \textbf{(a)} 1-D profile of star stacks of UVIT-N242W (blue-cyan), HST-F160W (orange), and ISAAC-Ks bands (green), with their peak values scaled to one for comparison. \textbf{(b)} The NUV model PSF (blue-cyan) with PSF matching kernals obtained for matching F160W (yellow) and ISAAC-Ks with N242W (green)  \textbf{(c)} The curve of growth profiles of N242W, N242W psf matched F160W and ISAAC-Ks bands. \textbf{(d)} Surface brightness (SB) profiles of AUDFs\_N02766 in the HST-UV bands (F225W and F275W), PSF-matched to the N242W resolution, compared with the N242W SB profile. Shaded regions indicate uncertainties in the HST-UV profiles. All profiles are shown without correction for extinction or attenuation. }
    \label{fig:PSF_matching}
\end{figure*}

The PSF matching routine for HST and ISAAC images to the N242W is listed below.
\begin{itemize}
    \item We identified isolated stars in the N242W image and ISAAC bands to model their PSF with a central moffat and extended exponential component, whose 1D radial profiles are shown in figure \ref{fig:PSF_matching}.
    \item Obtained the HST-F160W psf provided from \cite{2016Momcheva_3dhst}. 
    \item Psf matching kernels were obtained with photutils psf matching routine, after scaling all psfs to the same pixel scale. 
    \item The PSF matching kernels were tested by convolving HST psf and ISAAC psfs models, and comparing their curve of growth profiles with the curve of growth profile of NUV PSF (figure \ref{fig:PSF_matching}).
    \item The HST\_F160W-N242W PSF matching kernel was applied to other HST bands, which were originally matched to F160W in \cite{Whittaker_2019ApJS..244...16W}. While we obtained individual matching kernals for ISAAC J, H and Ks bands.
\end{itemize}

\subsection*{Photometry and SED fitting}
We work on Image cutouts of size $\sim 32.1\arcsec\times 32.1\arcsec$ obtained around the galaxy coordinates in the UVIT- F154W and N242W bands($77\times 77 $ pixels) and deep archival Optical and IR data, which are Issac- J, H, Ks bands($215\times 215$ pixels), Irac- 1,2,3,4 bands($107\times 107$ pixels) and HST(convolved to F160W psf)- F336W, F435W, F606W, F775W, F814W, F850LP, F098M, F125W, F140W, F160W bands($539\times 539$ pixels). All cutouts are shown in Figure \ref{fig:rgb_XUV_AUDF-X01}.

To ensure consistent flux measurements across bands, we PSF-matched the HST and ISAAC images to the UVIT N242W resolution. This was done using the PSF matching routine from Photutils \citep{Photutils_2022zndo....596036B}, with full details provided in Appendix \ref{appendix:psf_match}. The F154W band has a curve of growth (COG) that closely matches that of N242W beyond $r > 2\arcsec$, with deviations under 1$\%$ \citep{kanak_saha_AUDFCat}, and thus was not further adjusted. The IRAC bands, particularly IRAC-2, 3, and 4, have significantly broader PSFs than UVIT. PSF matching to IRAC resolution would risk contamination from nearby sources; therefore, we used aperture-corrected fluxes directly from the observed images for these bands.

We define a visually selected circular aperture with a radius of 4.58$\arcsec$ centered on the galaxy’s bulge, encompassing nearly all flux within the optical extent and avoiding contamination from nearby sources. This aperture is applied uniformly across all bands (F154W, N242W, PSF-matched HST and ISAAC images, and IRAC bands). Fluxes and associated uncertainties are measured using the corresponding weight maps, which are PSF-convolved for the HST and ISAAC images. For UVIT bands, flux uncertainties are computed as the Poissonian standard deviation of the total flux and background within the aperture, given by:

\begin{equation}
    \Delta f = \frac{\sqrt{(f+b\times A)\times t_{exposure}}}{t_{exposure}},
    \label{eqn:Pois_Error}
\end{equation}

\noindent where f is the calculated flux of the galaxy within the aperture, b is the mean background per pixel, A is the aperture area in pixels and $t_{exposure}$ is the total exposure time of the image. 

Aperture correction factors were calculated for the used aperture from the PSF of all bands and multiplied by the aperture flux and aperture flux errors of the galaxy.The observed magnitudes are presented in the Table \ref{tab:Photometry}.

\begin{table*}
\begin{tabular}{lllll}
\hline
\textbf{Telescopes} & \multicolumn{4}{c}{\textbf{Filters and Integrated Magnitudes (Observed)}} \\ \hline
\textbf{UVIT} & N242W & F154W &  &  \\
 & 22.19 $\pm$ 0.02 & 23.62 $\pm$ 0.08 &  &  \\ \hline
\textbf{HST} & F160W & F140W & F125W & F098W \\
 & 19.09$\pm$0.002 & 19.25 $\pm$ 0.006 & 19.34$\pm$ 0.002 & 19.65 $\pm$ 0.003 \\
 & F850LP & F814W & F775W & F606W \\
 & 19.82 $\pm$ 0.003 & 20.03 $\pm$  0.003 & 20.12 $\pm$ 0.004 & 20.98 $\pm$ 0.005 \\
 & F435W & F336W & F275W & F225W \\
 & 21.94 $\pm$  0.008 & 21.67 $\pm$ 0.04 & 22.12$\pm$0.04 & 22.03$\pm$0.04  \\ \hline
\textbf{SPITZER} & IRAC-1 & IRAC-2 & IRAC-3 & IRAC-4 \\
 & 18.80 $\pm$ 0.001 & 19.21 $\pm$ 0.002 & 19.20 $\pm$ 0.003 & 19.34 $\pm$ 0.004 \\ \hline
\textbf{VLT-ISAAC} & H & J & Ks &  \\
 & 19.62 $\pm$ 0.02 & 19.28 $\pm$ 0.01 & 18.98$\pm$ 0.02 &  \\ \hline
\end{tabular}
\caption{\textbf{Photometry of AUDFs\_N02766 across all the bands shown in Figure \ref{fig:rgb_XUV_AUDF-X01}. }}
\label{tab:Photometry}
\end{table*}

To construct the broadband spectral energy distribution (SED) of the galaxy from the far-UV to the infrared (1300\,\AA\ to 9.5\,\textmu m), we use the SED fitting code \textsc{CIGALE} \citep{Cigale}. Observed fluxes were first converted to milliJansky and corrected for foreground extinction at the effective wavelength of each filter.

The following \textsc{CIGALE} modules were used:
\begin{itemize}
    \item \texttt{sfh2exp}: double exponential star formation history,
    \item \texttt{bc03}: \citep{2003BC03} single stellar population models,
    \item \texttt{nebular}: \citep{2011Nebular} nebular continuum and line emission,
    \item \texttt{dustatt\_modified\_starburst}: modified Calzetti (2000) \citep{Calzetti2000} attenuation law,
    \item \texttt{dl2014}: \citep{2014Draine_dust} dust emission model.
\end{itemize}

We explored various dust attenuation laws corresponding to the Milky Way (MW), Large Magellanic Cloud (LMC), and Small Magellanic Cloud (SMC). While all yielded similar star formation histories, the best-fitting model (reduced $\chi^2 = 1.2$) was obtained using Milky Way attenuation parameters. The resulting SED fit yields a stellar mass of $1.083 \times 10^{11}\ M_\odot$, a young stellar component of $2.64 \times 10^8\ M_\odot$, and a star formation rate (SFR) of $28.1 \pm 1.4\ M_\odot\,\text{yr}^{-1}$. The best-fit color excess is $E(B - V) = 0.132$. 

\section{Masking of nearby sources in N242W profile}
\label{Appendix:masking}
\begin{figure*}
    \includegraphics[width=0.7\textwidth]{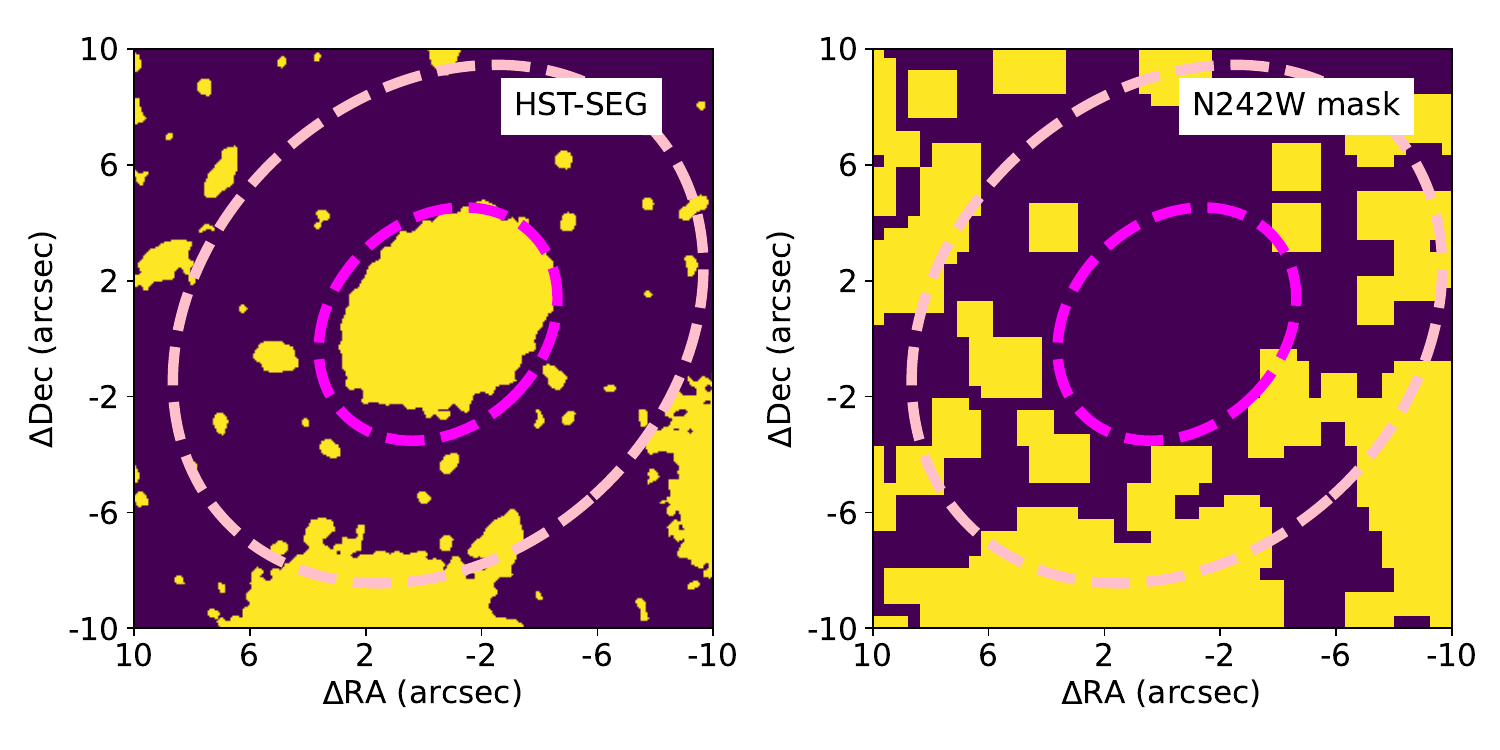}
    \caption{\textbf{(a)} The HST- Optical Segmentation map 
    made from the deep detection image \citep{Whittaker_2019ApJS..244...16W}.
    \textbf{(b)} Mask prepared to 
}
\label{fig:masking}
\end{figure*}

To mask neighbouring sources in the N242W image, we use the HST segmentation map from \citet{Whittaker_2019ApJS..244...16W}, which provides the deepest optical coverage of this field. We identify the segmentation ID corresponding to AUDFs\_N02766 and set it to zero to exclude the target from the mask. The resulting segmentation map is then resampled and convolved with a $2''\times2''$ box kernel to construct a mask suitable for removing contaminating sources in the UVIT N242W band. The HST segmentation and the corresponding N242W mask are shown in figure \ref{fig:masking}.

\section{Surface Brightness Corrections}
\label{Appendix:Surf_bright}
To obtain the rest-frame FUV and $K_s$-band surface brightness (SB) profiles, we apply cosmological surface brightness dimming, internal dust correction, and Galactic foreground extinction.

The cosmological dimming correction follows the relation from \citet{Hogg_Kcorr}:

\begin{align}
f_{\nu}(\nu_o) &= \frac{L_{\nu}(\nu_{e})(1+z)}{4\pi D_L^2}, \
\implies \mu(\nu_e) &= \mu(\nu_o) - 2.5\log(1+z)^3,
\end{align}

where $D_L$ and $D_A$ are the luminosity and angular diameter distances ($D_L=D_A(1+z)^2$), $L_\nu$ and $f_\nu$ are the luminosity and flux densities, $\nu_e$ and $\nu_o$ are emitted and observed frequencies, and $\mu$ is the SB in mag arcsec$^{-2}$, given by $\mu=ZP - 2.5\log(f/a)$, with $ZP$ the filter zeropoint and $a$ the aperture area in arcsec$^2$.

For our galaxy at $z=0.67$, the cosmological correction factor is $A_{\mathrm{cosmo}} = -2.5\log(1.67)^3 = -1.670$.

The internal dust correction is derived from the best-fit SED, giving $E(B-V)$ consistent with $A_{\mathrm{dust}}=-1.227$ at the rest-frame FUV wavelength ($\lambda=242/1.67=145$ nm) using the \citet{Calzetti2000} dust attenuation law. This correction is negligible in the $K_s$ band.

Foreground Galactic extinction values are taken from the NED extinction calculator, yielding $A_{\mathrm{fore}}=-0.054$ for N242W (negligible for IRAC-1).

The total correction for the rest-frame FUV SB profile is thus:

\begin{align}
\mu_{FUV} (\lambda_e=145~\mathrm{nm}) &= \mu (\lambda_o=242~\mathrm{nm}) + A_{\mathrm{cosmo}}+A_{\mathrm{dust}}+A_{\mathrm{fore}} \\
&= \mu (\lambda_o=242~\mathrm{nm}) - 2.951.
\end{align}

For the rest-frame $K_s$ band ($\lambda_e=2.2~\mu$m) derived from IRAC-1 ($\lambda_o=3.6~\mu$m), only the cosmological correction is significant:

\begin{align}
\mu_{K_s} (\lambda_e=2.2~\mu\mathrm{m}) &= \mu (\lambda_o=3.6~\mu\mathrm{m}) -1.670.
\end{align}

Thus, the N242W SB profile requires a total correction of $\approx3$ magnitudes to obtain the rest-frame FUV SB profile, and the IRAC-1 SB profile requires $\approx1.67$ magnitudes correction to obtain the rest frame Ks profile.

\label{lastpage}
\end{document}